\documentclass[a4paper,10pt,reqno,onesided]{amsart}

\usepackage{amsmath,amsfonts,amsthm,amssymb,dsfont,mathabx}
\usepackage[OT4]{fontenc}
\usepackage{enumerate}
\usepackage{mathrsfs,mathtools}
\usepackage{enumerate,enumitem, xspace}
\usepackage[pdftex]{graphicx}
\usepackage{color}
\usepackage{float}
\usepackage[center, font=small,labelfont=sc,labelsep=none]{caption}
\usepackage{euscript}
\usepackage{setspace}
\usepackage{fancyhdr}
\usepackage{url}
\usepackage{algorithm} 
\usepackage{algpseudocode} 
\algrenewcommand{\algorithmiccomment}[1]{// \textit{#1}}
\usepackage{thmtools}
\usepackage[colorlinks=true,allcolors=blue,backref=page]{hyperref}
\usepackage[noabbrev,capitalize,nameinlink]{cleveref}

\definecolor{darkred}{RGB}{200,0,0}
\hypersetup{colorlinks={true},linkcolor={darkred},citecolor={darkred}}

\usepackage[pagewise]{lineno}

\long\def\symbolfootnote[#1]#2{\begingroup
\def\thefootnote{\fnsymbol{footnote}}\footnote[#1]{#2}\endgroup}

\newtheorem*{question*}{Question}
\theoremstyle{remark}

\newtheorem*{claim*}{Claim}

\newtheorem*{remark*}{Remark}
\newtheorem*{example*}{Example}

\newtheorem*{fact*}{Fact}

\newtheoremstyle{named}{}{}{\itshape}{}{\bfseries}{.}{.5em}{\thmnote{#3}}
\theoremstyle{named}
\newtheorem*{namedtheorem}{Theorem}

\theoremstyle{plain}

\theoremstyle{definition}

\newcommand{\executeiffilenewer}[3]{%
\ifnum\pdfstrcmp{\pdffilemoddate{#1}}%
{\pdffilemoddate{#2}}>0%
{\immediate\write18{#3}}\fi%
}

\begin{document}

\title{on the non-locality of edge insertions}

\author[F.~Brunck]{Florestan Brunck}

\address{Institute of Science and Technology Austria\\
 Am Campus 1, 3400, Klosterneuburg, Lower Austria, Austria}
\email{florestan.brunck@mail.mcgill.ca}

\setlength{\headheight}{2em}
\setlength{\footskip}{2.5em}
\fancyhf{}
\fancyfoot[C]{\thepage}
\fancyhead[CE]{\textsc{on the non-locality of edge insertions}}
\fancyhead[CO]{\textsc{f.~brunck}}
\pagestyle{fancy}

\maketitle

\begin{abstract}
We challenge the idea that edge insertions are local improvement operations and show that the edge-insertion algorithm must sometimes insert an edge between vertices that are at the farthest combinatorial distance apart, and that this edge must also cross linearly many edges of the triangulation for the algorithm to escape a local optimum and return the optimal triangulation.
\end{abstract}

\section{introduction}

The edge insertion algorithm introduced by Edelsbrunner and al \cite{edge-insertion,min-max} provides an escape from the local optima where  regular edge flipping algorithms for triangulations tend to get stuck. It generalises the usual elementary diagonal flip in quadrilaterals to the more powerful operation of inserting an edge between any two vertices in the triangulation, followed by the retriangulation of the polygonal region left after removing all edges of the original triangulation which crossed the inserted edge. It currently offers the best polynomial algorithms for a number of optimisation problems, such as the min-max angle problem (\cite{min-max}). While flips are inherently local elementary operations, edge insertions are not. It is then natural to wonder exactly how non-local edge insertions have to be to guarantee optimisation. Perhaps surprisingly - especially in contrast with the closely related max-min problem where the Delaunay triangulation can be reached via flips - we show that edge insertions cannot be constrained to be local operations at all when trying to optimise triangulations for the min-max angle problem. In particular, we give an example where the edge insertion algorithm must insert an edge between vertices that are at the farthest combinatorial distance apart, and that this edge must also cross linearly many edges for the edge insertion algorithm to escape a local optimum and return the min-max angle triangulation.

\begin{namedtheorem}[Proposition]
\label{prop}
Edges-insertions sometimes cannot be local operations. For any integer $n\in\mathbb N$, there exists a triangulation $\mathcal T_n$ with $4n+5$ edges and combinatorial diameter $n+2$, such that, in order to successfully return the min-max angle triangulation, the edge insertion algorithm must insert an edge between vertices which are at combinatorial distance $n+2$. Furthermore, that edge intersects $2n$ edges of~$\mathcal T_n$.
\end{namedtheorem}

\section{background - the edge insertion algorithm}

Recall that a \emph{triangulation} of a finite set of points $\mathcal S$ in the Euclidean plane is a maximally connected, straight-line planar graph with set of vertices $\mathcal S$. We denote by $\mu$ the function that maps a triangle $t$ in $\mathcal T$ to the value of its maximum angle. The example we provide concerns the \emph{min-max angle problem}, namely the task of finding the triangulation of a fixed point set in the plane which minimises the maximum value of $\mu$ over all triangles $t$ in the triangulation. The \emph{measure} $\mu(\mathcal T)$ of a triangulation $\mathcal T$ is defined as the quantity $\operatorname{max}\{\mu(t)\> | \> t \text{ a triangle in } \mathcal T\}$. If $\mathcal T$ and $\mathcal T'$ are two triangulations of the same point set, we say that $\mathcal T$ is an improvement of $\mathcal T'$ and write $\mathcal T \prec \mathcal T'$ if $\mu(\mathcal T)<\mu(\mathcal T')$ or if $\mu(\mathcal T)=\mu(\mathcal T')$ and the set of triangles $t$ in $\mathcal T$ such that $\mu(t)=\mu(\mathcal T)$ is a strict subset of the set of such triangles in $\mathcal T'$. A triangulation $\mathcal T$ is \emph{optimal} if there is no improvement of $\mathcal T$. 

\begin{figure}[H]\centering
  \includegraphics[page=4]{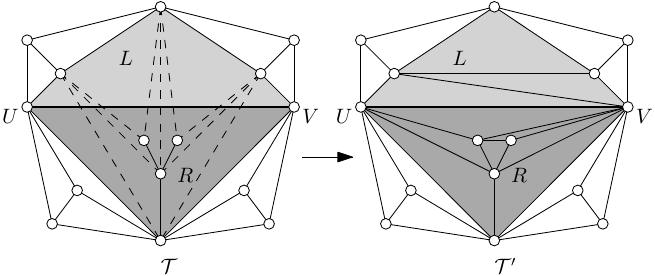}
  \caption{: The edge insertion of an edge $UV$ (in bold) in a triangulation $\mathcal T$ of a planar point set. The original edges of $\mathcal T$ which intersect $UV$ are shown in dashed lines on the leftmost diagram while the optimally retriangulated ones are shown on the rightmost diagram. The affected regions $L$ and $R$  are highlighted in light and dark grey.}
  \label{fig:edge-insertion}
\end{figure}

Given a triangulation $\mathcal T$ of a point set $\mathcal S$ in the plane, the \emph{edge insertion} $UV$, $U,V\in \mathcal S$ is the following procedure (\cref{fig:edge-insertion}):

\begin{algorithm}[H]
	\caption*{\textbf{Edge Insertion}\boldmath $\pmb(\mathcal T, UV \pmb )$} 
	\begin{algorithmic}
        \State $\mathcal T' \longleftarrow$ $\mathcal T$
        \State Add the edge $UV$ and remove from $\mathcal T'$ all edges that intersect $UV$
        \State \algorithmiccomment{This creates two polygonal regions $L$ and $R$ on either side of the inserted edge}
        \State Retriangulate $L$ and $R$ optimally
        \State \algorithmiccomment{In the simplest version this is done by dynamic programming}
        \State Return $\mathcal T'$
	\end{algorithmic} 
\end{algorithm}

The \emph{edge insertion algorithm} is then the following:

\begin{algorithm}[H]
	\caption*{\textbf{Edge Insertion Algorithm}\boldmath $\pmb(\mathcal S \pmb )$} 
	\begin{algorithmic}
        \State \algorithmiccomment{Input: $\mathcal S$ a planar point set}
        \State \algorithmiccomment{Output: an optimal triangulation $\mathcal T_{\text{opt}}$ for $\mathcal S$}
        \State Construct an arbitrary triangulation $\mathcal T$ of $\mathcal S$ 
        
		\Repeat { $\mathcal T' \longleftarrow \mathcal T$}
            \For {all edges $UV$ with $U\neq V\in\mathcal S$}
                \State $\mathcal T''\coloneqq \operatorname{Edge\>Insertion}(\mathcal T, UV)$
                \If {$\mathcal T'' \prec \mathcal T$} {$\mathcal T \longleftarrow$ $\mathcal T''$ and \textbf{exit the for-loop}}
                \EndIf 
            \EndFor
        \Until $\mathcal T=\mathcal T'$
	\end{algorithmic} 
\end{algorithm}

The correctness of the above algorithm was proven in \cite{min-max}. By being more parsimonious in which candidate edge insertions should be considered, and getting rid of dynamic programming for the retriangulation step in favour of a smarter iterated ear-removal process, the following running time was obtained:

\begin{namedtheorem}[Theorem (Edelsbrunner, Waupotitsch and Tan)]
    Given a finite set of points $\mathcal S$ in the plane, the edge insertion algorithm returns the optimal min-max angle triangulation in time $O(|\mathcal S|^2 \log |\mathcal S|)$.
\end{namedtheorem}

\section{the manta ray triangulation}

In this section, we describe the construction of our pathological triangulation and prove our \nameref{prop}.
\begin{proof}[Proof of the \nameref{prop}.]\noindent
We describe the construction for $\mathcal T_n$, as seen on \cref{fig:example-1}. Starting with an isosceles triangle $OAB$, consider the ray $r_A$ (resp. $r_B$) starting at $A$ (resp. $B$) and meeting the line $OA$ (resp. $OB$) with a clockwise (resp. counterclockwise) angle of $\theta$. The direction of the rays is chosen such that, for any point $R$ lying on $r_A$ (resp. $r_B$), the triangle $ARB$ is direct. Let us fix $n\in\mathbb N$. We inductively construct points $A_0,A_1,\ldots, A_n$ and $B_0,B_1, \ldots, B_n$ in the following way:

\begin{itemize}
\item $A_0=A$, $B_0=B$.
\item For all $0 \leq i\leq n$, $|A_iA_{i+1}|=|B_iB_{i+1}|=|A_iB_i|$.
\end{itemize}

\begin{figure}[H]\centering
  \includegraphics[page=2]{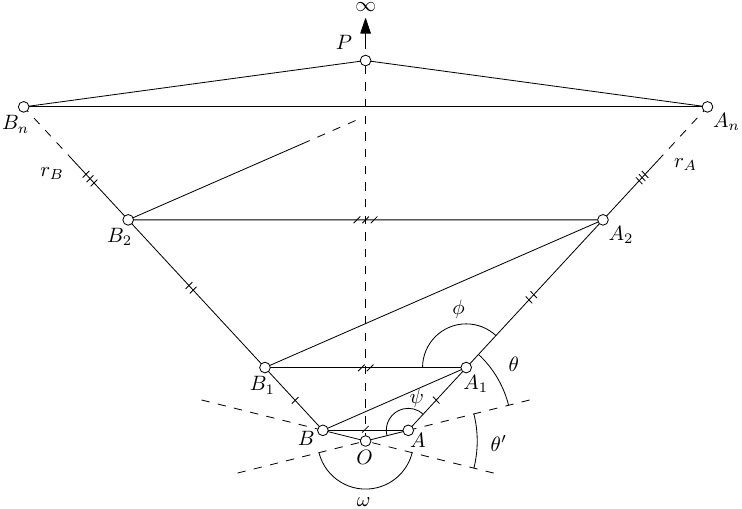}
  \caption{: The ``manta ray'' triangulation $\mathcal T_n$.}
  \label{fig:example-1}
\end{figure}

We additionally define a point $P$, lying on the perpendicular bisector of $A$ and $B$, ``far away'' from $A$ (exactly how far will be made precise shortly). We can then consider the triangulation containing the triangles $OAB$, $A_nPB_n$, as well as the edges $(A_i,B_i)$, $(A_{i+1},B_i)$, $(A_i,A_{i+1})$ and $(B_i,B_{i+1})$, for $0\leq i\leq n-1$. We readily check that this triangulation has $4n+5$ edges and combinatorial diameter $n+2$. \\
Let us introduce some additional notation for the angles. By construction, for $1\leq i \leq n-1$, the angles $\angle A_{i+1}A_iB_i$ are all equal, let us call this angle $\phi$. The triangle angle at $O$ in $OAB$ will be denoted by $\omega$ and the angle $\angle A_1BO$ by $\psi$. Call $\theta$ the complementary angle $\pi - \psi$. We also denote by $\alpha$ the largest angle between the line $OA$ and the perpendicular to $AB$ passing through $A$ (see \cref{fig:example-2}). By construction, for all $1\leq i \leq n$, the angle between the ray $r_A$ and the perpendicular to the line $AB$ passing through $A_i$ does not depend on $i$. We denote that angle by $\beta$. Note that this triangulation has two degrees of freedom and can be fully parametrised by $\omega$ and $\theta$, so that we may refer to it as $\mathcal T_n(\omega,\theta)$.

The proof of the proposition follows from the following claim after taking $\mathcal T_n\coloneqq \mathcal T_n(\omega_0,\theta_0)$:

\begin{claim*}
\noindent There exists a value of $\omega_0$ and $\theta_0$ such that the following inequalities hold in the triangulation $\mathcal T_n(\omega_0,\theta_0)$:
\begin{enumerate}[label=(\roman*)]
    \item $\phi<\omega$
    \item $\psi>\omega$
    \item $\alpha < \omega$
    \item $\beta < \omega$
\end{enumerate}

\end{claim*}
Inequality $(i)$ shows that $\omega$ is the largest angle in $\mathcal T_n$. Inequality $(ii)$ shows that any edge insertion that does not use the edge $OP$ does not yield an improvement of $\mathcal T_n$. Indeed, any edge $OX$ with $X=A_i$ or $X=B_i$, $1 \leq i \leq n$, forces the existence of the triangle $OAA_1$ or $OBB_1$ in the triangulation. Finally, inequalities $(iii)$ and $(iv)$ show that inserting the edge $OP$ and re-triangulating by joining $P$ to all the other vertices indeed yields an improvement of $\mathcal T_n$ (in fact, the min-max angle triangulation itself). Indeed, we can select $P$ far enough towards infinity so that these last two inequalities are not only true for $\alpha$ and $\beta$ but also for the angles $\angle OAP$ and $\angle AA_1P$.

\begin{figure}[H]\centering
  \includegraphics[page=3]{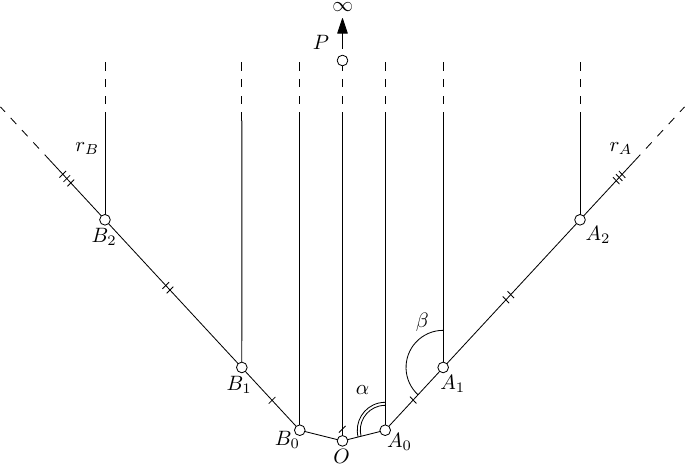}
  \caption{: The optimal triangulation minimising the maximum angle.}
  \label{fig:example-2}
\end{figure}

\begin{proof}[Proof of the claim.]
We first eliminate a degree of freedom by adding the constraint $\theta = \frac{2}{3}\theta'$ (see \cref{fig:example-1}). Through straightforward angle-chasing, one can then check that enforcing $\omega>\frac{10}{13}\pi$ (and that $P$ is far enough to infinity to guarantee $\angle A_nPB_n < \omega$) gives us suitable values for the claim and we need only fix any $\omega_0$ in that range (which then also fixes the value of $\theta_0)$.
\end{proof}

\end{proof}

\begin{remark*}
If we want the points to be in general position, we can simply perturb the $A_i$ (resp. $B_i$) to lie on a concave curve by nudging $A_1$ (resp $B_1$) slightly towards the inside of the basket (to ensure the triangles $OAA_1$, $OBB_1$ are still needed) and nudging $A_{i+1}$ so that the segment $A_{i-1}A_{i+1}$ lies outside of the basket (similarly for each $B_i$).
\end{remark*} 

\section*{acknowledgements}

I am indebted to insightful discussions with Herbert Edelsbrunner, which sparked my curiosity and prompted this observation.

\bibliography{references.bib}
\bibliographystyle{plain}

\end{document}